\documentclass[preprint, superscriptaddress,onecolumn,a4paper]{revtex4}%

\usepackage{bbm}
\usepackage{amsfonts}
\usepackage{amssymb}
\usepackage{graphicx}
\usepackage{subfigure}
\usepackage{savesym}
\usepackage{amsmath}
\usepackage{txfonts}
\usepackage{multirow}
\usepackage{epstopdf}
\usepackage{color}

\newcommand{\ket}[1]{\vert#1\rangle}

\usepackage{pbox}
\usepackage{amsmath}

\definecolor{VK}{RGB}{0, 95, 0}

\begin{document}

\title{Efficient polarisation-preserving frequency conversion from a trapped-ion-compatible wavelength to the telecom C band}

%

\author{V. Krutyanskiy}

	\affiliation{Institut f\"ur Quantenoptik und Quanteninformation,\\
	\"Osterreichische Akademie der Wissenschaften, Technikerstr. 21A, 6020 Innsbruck,
	Austria}

\author{M. Meraner}

	\affiliation{Institut f\"ur Quantenoptik und Quanteninformation,\\
	\"Osterreichische Akademie der Wissenschaften, Technikerstr. 21A, 6020 Innsbruck,
	Austria}
	\affiliation{
	Institut f\"ur Experimentalphysik, Universit\"at Innsbruck,
	Technikerstr. 25, 6020 Innsbruck, Austria}
	
\author{J. Schupp}

	\affiliation{Institut f\"ur Quantenoptik und Quanteninformation,\\
	\"Osterreichische Akademie der Wissenschaften, Technikerstr. 21A, 6020 Innsbruck,
	Austria}
	\affiliation{
	Institut f\"ur Experimentalphysik, Universit\"at Innsbruck,
	Technikerstr. 25, 6020 Innsbruck, Austria}

\author{B. P. Lanyon}
\email{ben.lanyon@uibk.ac.at}

	\affiliation{Institut f\"ur Quantenoptik und Quanteninformation,\\
	\"Osterreichische Akademie der Wissenschaften, Technikerstr. 21A, 6020 Innsbruck,
	Austria}

\begin{abstract}
We demonstrate polarisation-preserving frequency conversion of single-photon-level light at 854 nm, resonant with a trapped-ion transition and qubit, to the 1550-nm telecom C band.
A total photon in / fiber-coupled photon out efficiency of $\sim$~30~\% is achieved, for a free-running photon noise rate of $\sim$~60~Hz.
This performance would enable telecom conversion of 854-nm polarisation qubits, produced in existing trapped-ion systems, with a signal-to-noise ratio greater than 1. 
In combination with near-future trapped-ion systems, our converter would enable the observation of entanglement between an ion and a photon that has travelled more than 100 km in optical fiber: three orders of magnitude further than the state-of-the-art.  

\end{abstract}

\maketitle
%
%
\section{Introduction}

There is a current multi-disciplinary research initiative to develop light-matter quantum networks \cite{Kimble:2008fk};  remote nodes, consisting of precisely-controllable quantum matter in which quantum information is stored and processed, that are interconnected with quantum light, such as single photons. These networks, envisaged over distances from table-top to intercontinental, could enable a  range of new science and technology, including scalable quantum computing \cite{PhysRevA.89.022317}, secure communication \cite{RevModPhys.74.145} and enhanced sensing \cite{sensingnetwork}. 

Given the great success in encoding, manipulating, storing and reading-out quantum information in their electronic states, trapped atomic ions represent a powerful platform with which to build, or integrate into, the nodes of quantum networks \cite{RevModPhys.82.1209, PhysRevA.79.042340}. Indeed, an elementary quantum network consisting of ions in two traps a few meters apart, has been entangled via travelling ultraviolet photons \cite{Moehring:2007qf}. A challenge is that most readily-accessible photonic transitions in trapped ions lie at wavelengths that suffer significant absorption loss in materials for manipulating and guiding light, thereby limiting the internode networking distance. Another challenge is that ionic transitions are fixed and narrowband, such that, except in rare cases \cite{PhysRevA.80.062330}, they cannot be interfaced with other examples of quantum matter to enable new ion-hybrid quantum systems \cite{Kurizki31032015}. 

The aforementioned challenges could be overcome using quantum frequency conversion (QFC) \cite{PhysRevLett.68.2153, Tanzilli:2005ad}; a nonlinear optical process in which a photon of one frequency is converted to another, whilst  preserving all the quantum and classical photon properties. QFC of single photons has recently been studied in a variety of contexts \cite{Pelc:11, 13NaPh_Lavoie, 16PRL_Gaeta_Ramsey, Ikuta:13, Zhou:16, Alibart:1}  and is typically achieved using three-wave mixing in a second-order non-linear ($\chi^{(2)}$)  crystal. It has been shown that QFC can preserve a broad range of photon properties, including first- and second-order coherence, and pre-existing photon-photon entanglement \cite{Tanzilli:2005ad, PhysRevA.85.013845, Lenhard:17}. QFC could therefore act as a quantum photonic adapter for trapped ions, allowing their high-energy photonic transitions to be interfaced with the lower-energy photons better suited for long-distance travel through optical fibers, or with other forms of quantum matter. 

Interfacing trapped ions with the telecom wavelengths of 1310  nm (O band) or 1550 nm (C band) is particularly appealing: these wavelengths suffer minimal transmission losses (0.32 and 0.18 dB/km, respectively) through optical fibers and a broad range of established technologies and infrastructure for their manipulation and transmission exist.  The telecom wavelengths are therefore an ideal choice for a universal standard for light-matter quantum networks, allowing similar and dissimilar quantum matter to interface over both short and long distances. 

Telecom frequency conversion of photons connected to several examples of quantum matter has recently been demonstrated, including quantum dots \cite{PhysRevLett.109.147404,Kambs:16,Pelc:12, Yu2015}, cold gas atomic ensembles \cite{Albrecht:2014kk, Farrera:16, Ikuta:16} and solid-state ensembles \cite{huguescrystal}. Applying QFC to trapped ions is challenging. The comparatively low rate and efficiency with which photons have been collected from / absorbed by an ion demands a highly efficient and low noise conversion process. Readily accessible photonic transitions in ions also lie in the ultraviolet or visible regime, which suffer high absorption and strong dispersion in nonlinear crystals. Furthermore, direct (single step)  conversion of those  photons to telecom in the so called `long pump wavelength regime' is not possible, leading to additional noise processes during conversion \cite{Pelc:11}. Nevertheless, significant progress has been made in overcoming these challenges \cite{Clark2012, lobino, PhysRevApplied.7.024021}. In \cite{PhysRevApplied.7.024021}, for example, the authors convert attenuated laser light at 369.5 nm (a transition in Yb$^+$) to 1311 nm, achieving a waveguide efficiency of $\sim$ 5 \% (including coupling losses) and a total efficiency for fiber-coupled output photons of 0.4 \%.  

In this paper, we present experiments that demonstrate photon conversion from $\lambda_{s}=$ 854 nm ($s=$ signal) to $\lambda_t=$ 1550 nm ($t=$ target, Telecom C band), via difference-frequency generation (DFG) in a waveguide-integrated $\chi^{(2)}$ crystal, using a strong pump laser at $\lambda_p=$ 1902~nm. The signal wavelength corresponds to the $P_{3/2}$ to $D_{5/2}$ dipole transition in singularly ionised atomic Ca$^+$, which can be efficiently collected from the ion in a cavity quantum electrodynamic (CQED) setting. 
Our experiments use laser light, resonant with the ionic transition and attenuated to the single-photon level. We refer the reader to work studying conversion from 854 nm to 1310~nm \cite{thesisLenhard}. 

The content of the paper is as follows. First, the importance of the 854-nm transition in Ca$^+$ is briefly discussed and motivated. Section \ref{scheme1} presents a scheme that enables telecom conversion of one polarisation component of an 854-nm photon. Here, the limits on the efficiency and photon noise are presented. Section \ref{scheme2} presents a scheme that preserves the polarisation during conversion, allowing translation of a polarisation qubit from 854 nm to 1550 nm with high fidelity. The achieved performance brings QFC experiments within reach of existing 854-nm trapped-ion photon sources. 
Section \ref{long-distance} discusses the potential for our conversion scheme, combined with future trapped-ion sources, to enable the observation of entanglement between an ion and a photon that has travelled more than 100 km of optical fiber. Section \ref{conc} presents our discussion and conclusion.

\subsection{The 854 nm transition in Ca$^+$} \label{854}
From the perspective of achieving efficient, low-noise photon frequency conversion from a trapped-ion wavelength to telecom, the 854-nm transition in Ca$^+$ represents an ideal choice. Transmission losses at this wavelength are low in non-linear conversion crystals and the required poling period to overcome the phase mismatch can be precisely manufactured, allowing for efficient first-order quasi phase matching and long interaction lengths. Futhermore, single-step conversion to 1550 nm requires a pump laser in the so-called long-pump-wavelength regime: the 1902-nm pump photons have lower energy than the target 1550-nm wavelength, such that spontaneous parametric down conversion (SPDC) of the pump cannot produce noise photons at 1550 nm \cite{Pelc:11}. The threshold input photon wavelength for this condition is $\lambda>$ $1550/2~\mathrm{nm} =775$ nm ($1310/2~\mathrm{nm} =655$ nm), which can be found in very few ionic species and in each case is branching-ratio unfavored.  

From the perspective of the ion, the 854-nm transition is directly connected to a leading qubit transition \cite{schindlerCa}: emission of an 854-nm photon by Ca$^+$ leaves the ion's valence electron in the metastable excited state $\ket{D_{5/2}}$ of the well known 729-nm optical qubit.  A broad array of techniques are readily available for manipulating the quantum state on this optical qubit transition \cite{schindlerCa}. It is a challenge to efficiently collect 854-nm photons from the ion (or have them absorbed by the ion): in free space, resonant excitation to the excited $\ket{P_{3/2}}$ leads to the emission of an 854-nm photon in only $\sim$1/17 of cases \cite{Gerritsma2008}, in most cases a $393$-nm photon is emitted. However, an optical cavity around the ion can enhance emission on this transition \cite{Keller:2004fe, Stute:2012fk}. In such a CQED system, both near-maximal entanglement between a travelling $854$-nm polarisation qubit and the ion qubit \cite{Stute:2012fk}, and state mapping from ion qubit to photon \cite{statetransfer} have been achieved with high fidelity. 
The bandwidth of an 854-nm photon in a CQED setup can be significantly narrower than atomic transition linewidth in free space (23 MHz), as the photon leaks slowly out of a high-finesse cavity (e.g. $\sim$ 50 kHz \cite{Stute:2012fk}). 
Such narrowband photons present challenges and opportunities for low noise QFC. On the one hand, noise photons introduced by the QFC process will be integrated over the temporally long photon wave packets (many microseconds). On the other hand, the narrowband photons can be strongly spectrally filtered to reduce such noise, without compromising transmission.

\section{Polarisation-dependent conversion scheme}  \label{scheme1}

\subsection{Experimental details} \label{expdetails}
We perform frequency conversion by DFG using the $\chi^{(2)}$ nonlinearity in a LiNbO$_3$ waveguide-integrated chip. Each chip is 48~mm long and contains ridge waveguides (LiNbO$_3$ layer on LiTaO$_3$ substrate) milled out along its length, with dimensions of approximately 11.0 $\mu$m by 12.1 $\mu$m (fabricated by NTT electronics). To achieve first-order quasi-phase matching, the guides are poled with a period of approximately 22 $\mu$m. While each chip  has 12 individual waveguides, a single guide in any one chip is used for each experiment. The waveguides enable a continuous high spatial mode overlap between the three optical fields, are single mode for 1550 nm and 1902~nm, multimode for 854 nm and are anti-reflection coated for all those wavelengths on each facet ($R\leq1$\%). The conversion process is phase matched when all three optical fields have the same linear polarisation; the orthogonal polarisation is supported in the guide but remains unconverted. 

We now provide a summary of the basic experimental setup, detailed in Figure \ref{figure1}. For the pump we use a Tm-doped fiber laser (AdValue Photonics AP-SF1) and for the 854 nm input a diode laser (Toptica DLPRO), stabilised to within a few MHz of the ionic transition using a wavemeter lock (High Finesse WSU10). Both signal and pump are delivered to the photon conversion setup using polarisation-maintaining single-mode optical fiber, spatially overlapped using a dichroic mirror and free-space coupled into a ridge waveguide via an aspheric lens. A second asphere at the waveguide output collimates the output fields, which are then sent to various filtering and analysis stages. The chip is temperature-stabilised and a waveguide is chosen with a quasi-phase matching temperature of  38$^{\circ}$C. The spectral acceptance bandwidth of the phase matched conversion process centred at 854 nm is measured to be $\sim$  0.2 nm (82 GHz), which agrees with theoretical calculations based on refractive indices of bulk  $\mathrm{LiNbO_3}$ at the corresponding wavelengths \cite{Zelmon:97}. Note that this acceptance bandwidth for photon conversion is orders of magnitude broader than 854 nm photons from the ion, and does therefore not act as a filter. The temperature bandwidth (FWHM) of the phase matched conversion process is measured to be $2.4\pm0.2$ $^{\circ}$C. 

 \subsection{Results} \label{results1}
 
 \subsubsection{Efficiency.} 
 
 QFC in a material with a $\chi^{(2)}$ nonlinearity is a three-wave mixing process in which the quantum states of light can be coherently interchanged between two frequencies $\omega_1$ and $\omega_2$ via the interaction with a strong (undepleted) pump field at frequency $\omega_p$, where $\omega_1+\omega_p=\omega_2$. In our experiments, $\omega_1$ and $\omega_2$ are the frequencies of the 1550 nm and 854 nm photons, respectively. For QFC in a waveguide-integrated lossless material, and in the case of perfect phase matching, one can show that the efficiency of conversion for an interaction (waveguide) length $L$ is given by \cite{Roussev:04}  $\eta = N_2/N_1 =  \sin^2{(\sqrt{\eta_{nor}P_p}L)}$. Here, $N_2$ is the number of output 1550 nm photons, $N_1$ the number of input 854 nm photons, $P_p$ is the pump power. The normalised efficiency $\eta_{nor}$ depends on the material nonlinear strength and the three-wave mode-overlap integrals. 
For a given waveguide length $L$, there is a pump power $P_{max}$ that achieves complete conversion. For larger pump powers, conversion back to the initial frequency takes place.

\begin{figure}[th]
	\centering
  \includegraphics[width=0.5\textwidth, angle=0]{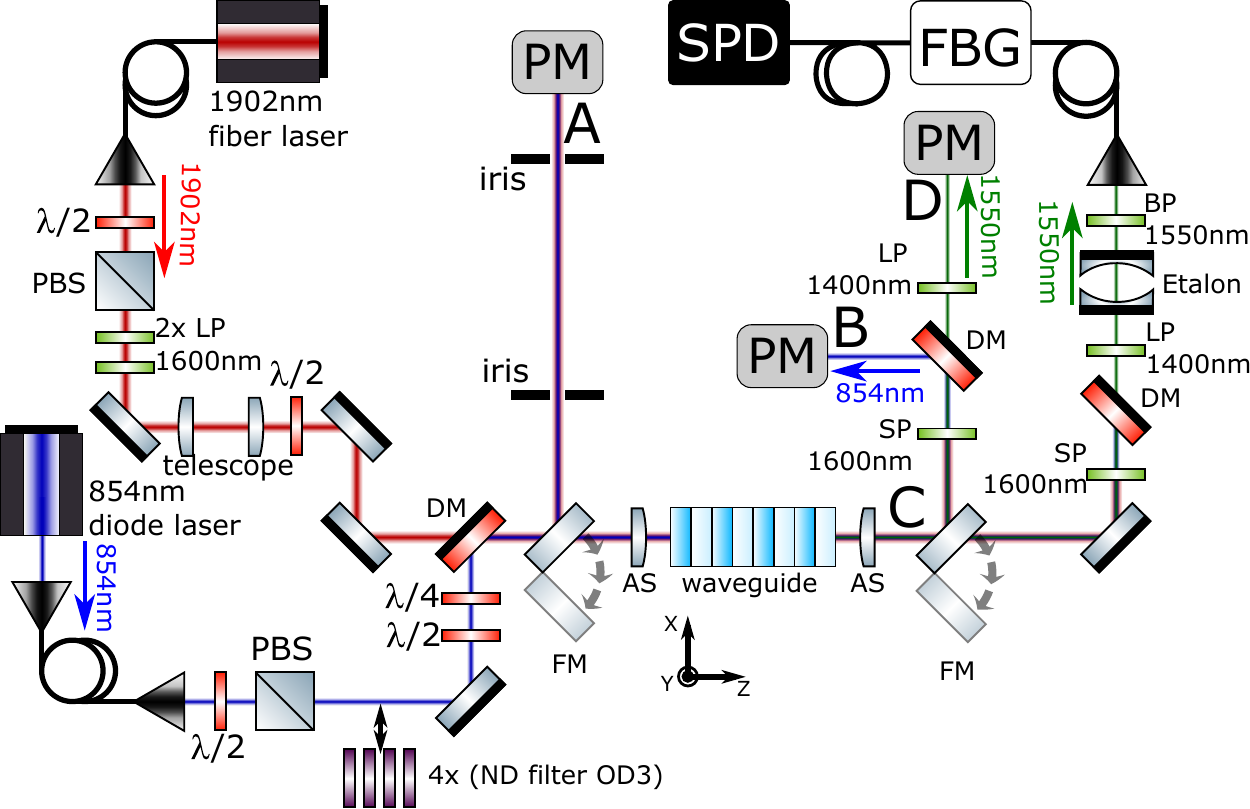}
	\caption{\textbf{Setup for polarisation-dependent frequency conversion from 854 nm to 1550 nm}. \textcolor{black}{LP 1600 nm -- longpass filters (Edmund Optics 84-680); DM -- dichroic mirror (Layertec 103472, highly reflective for 854 nm, transmissive for 1550 and 1902nm); FM -- gold flip mirror; AS -- 11mm asphere lenses (Thorlabs A220TM) positioned by XYZ translation stage; SP 1600 nm -- shortpass filter (Edmund Optics 84-656, OD5 at 1902 nm); LP 1400 -- longpass filter (Thorlabs FELH1400) prevents detecting second harmonic of pump; BP 1550 nm -- bandpass filter (Thorlabs FB-1550, 63\% transmission, OD5); Etalon -- air-spaced Fabry-P\'{e}rot cavity (SLS optics): 250 MHz linewidth, 12.5~GHz free spectrum range, peak transmission 95\% at 1550 nm, extinction $10^3$; FBG -- fiber Bragg grating filter (Advanced Optical Solutions): bandwidth 15 pm (2 GHz), transmission 60\% at 1550 nm, extinction  $10^3$; PM -- power meters; SPD -- avalanche photo diode based single telecom photon detector (see text).
}}
	\label{figure1}
\end{figure}

We begin by sending 315 $\mu$W of 854 nm laser light to the waveguide, optimising the in-coupling to maximise the population of the lowest order (fundamental) mode using a beam profiler at the output. For the analysis of the output fields,  optical power meters are used. The input 854 nm power is measured at point A, $P^{A}_{854}$, the output at point B, $P^{B}_{854}$ and the transmission at 854 nm is quantified as $T_{854}^{B/A}=P^{B}_{854}/P^{A}_{854}$. For zero pump power, we find $T_{854}^{B/A}= 0.73\pm{0.01}$.  
The known relevant optical losses at 854 nm between points A and B are: two uncoated aspheres (measured transmission $T_a=0.93$ each); shortpass filter (measured transmission $T_{sp}=0.96\pm{0.03}$); dichroic mirror (reflection $R_d=0.99$). 
When considering these losses we find a waveguide transmission of $T_{854}^{waveguide}= T_{854}^{B/A}/(T_{a}^2T_{sp}R_{d})=0.89\pm{0.04}$, that now contains losses only due to waveguide in-coupling and propagation. 
%
The input (output) 1902 nm pump power is measured at point A (C), yielding a pump transmission of $T_{1902}^{C/A}= 0.57\pm{0.01}$. 
 The output 1550 nm power $P^{D}_{1550}$ is measured at point D. 

\begin{figure}[ht]
	\centering
  \includegraphics[width=0.5\textwidth, angle=0]{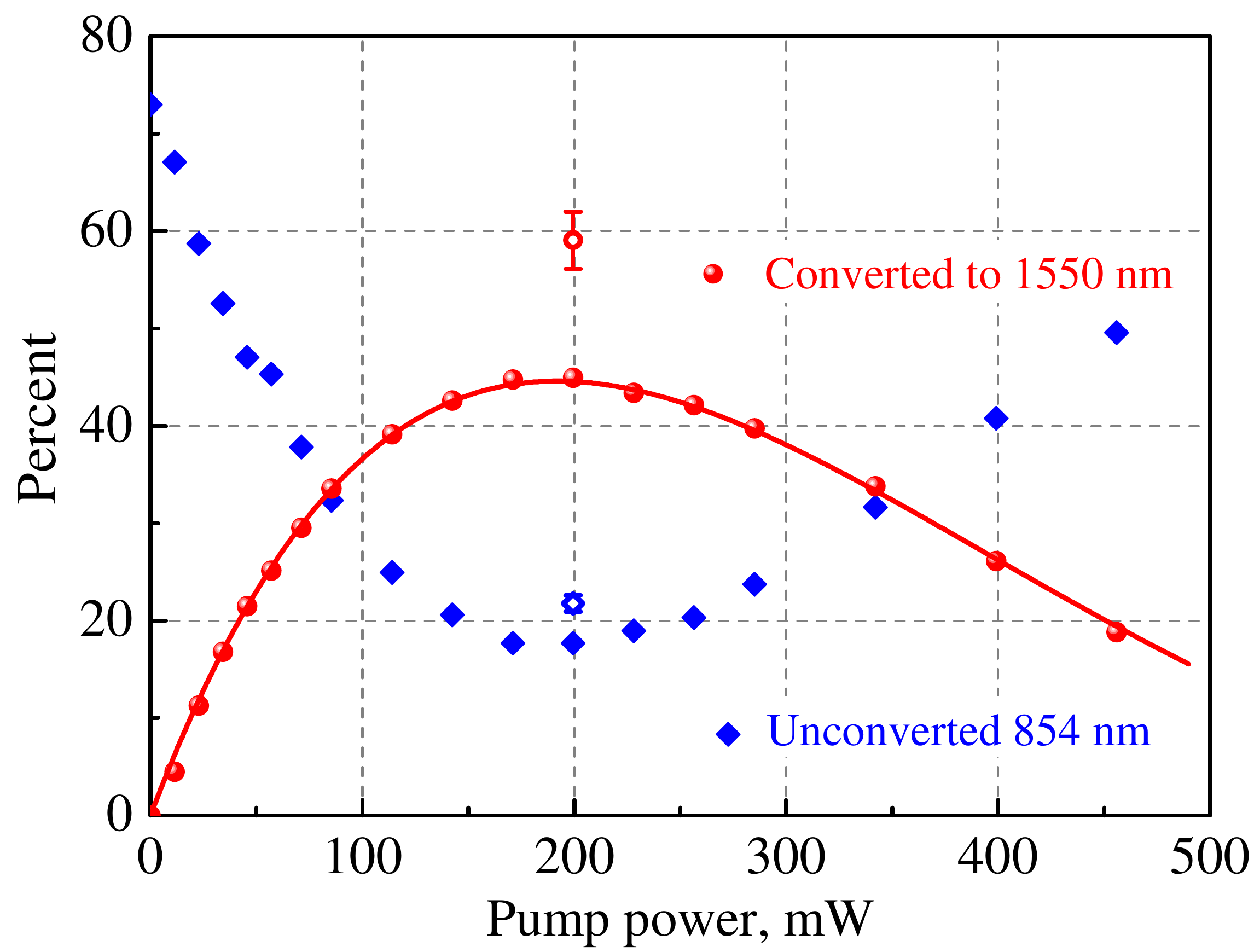}
	\caption{\textbf{Conversion efficiency, of 315 $\mu$W of 854 nm laser light to 1550 nm, as a function of pump power}. Solid red spheres: conversion efficiency to 1550 nm $\eta^{A/B}_{classical}$. Unfilled red circle: waveguide conversion efficiency to 1550 nm $\eta^{waveguide}_{classical}$, after accounting for known passive optical losses. 	
Solid red line: theoretical fit by the function $\eta = A\cdot\sin^2{(\sqrt{\eta_{nor}P_p}L)}$ with fitting parameters $A = 0.45$, $\eta_{nor} = 0.56\mathrm{W^{-1}cm^{-2}}$).
Filled blue diamonds: fraction of remaining 854 nm light $T_{854}^{B/A}$ after the waveguide.  Unfilled blue diamond: fraction of remaining 854 nm light after the waveguide, when accounting for known passive optical losses. Visible error bars stem from uncertainty in passive loss of optical elements. 
 } 
	\label{figure2}
\end{figure}

Figure \ref{figure2} presents the measured conversion efficiency, quantified by $\eta^{D/A}_{classical}=P^{D}_{1550}/P^{A}_{854} \times 1550/854$ (equivalent to ratio of photon numbers), as a function of output pump power $P^{C}_{1902}$. The depletion of the 854 nm signal is also shown, quantified by the transmission at 854 nm $T_{854}^{B/A}$. The results show a maximum conversion efficiency $0.46~\pm{0.01}$  for a pump power of 200 mW.

When removing the known relevant losses in optics around the waveguide  at 854 nm and 1550 nm, one obtains a maximum external classical waveguide conversion efficiency of $\eta^{waveguide}_{class}=0.59 \pm{0.03}$, which still contains losses due to in-coupling into the waveguide and waveguide propagation losses \footnote{$0.59 = 0.46/(0.93\cdot 0.93\cdot 0.96\cdot 0.94\cdot 0.995)$, where the denominator contains transmissions at the relevant wavelength for: in-coupling asphere; out-coupling asphere; shortpass filter; dichroic mirror; longpass filter.}.

Assuming that the waveguide transmission losses are equal at 854 nm and 1550 nm, 0.89$\pm{0.04}$ ($T_{854}^{waveguide}$) is the maximum conversion efficiency. The value that we obtain, $\eta^{waveguide}_{class}$, is $0.30\pm{0.05}$ lower than this maximum. Figure \ref{figure2} shows that, at the point of maximum conversion efficiency, a fraction 0.22$\pm{0.01}$ of unconverted 854 nm light remains at the waveguide output (when accounting for known passive optical losses). One sees, therefore, that the majority of `missing conversion efficiency' lies in unconverted 854 nm light.  

In the case of perfect phase matching and no losses, the DFG process for three optical modes in a waveguide allows for complete depletion (conversion) of the signal wave. However, the situation is more complicated when several spatial (or axial \cite{PhysRevA.86.033827}) modes are involved. Indeed, our waveguides are multimode at 854 nm and a fraction of 854 nm populates higher-order modes due to imperfect mode matching at the input. The different modes have different effective refractive indices than the fundamental mode and are not therefore simultaneously phase matched.  
As a consequence, the conversion efficiency of higher order modes is weak and they remain largely unconverted. 
Beam profile measurements of the unconverted 854 nm light exiting the waveguide reveal that it consists of spatial modes of much higher order than the fundamental.

We conclude from this experiment that a waveguide conversion efficiency of 0.59 $\pm{0.03}$  was achieved, limited by the unintentional excitation of higher-order waveguide modes at 854~nm. The use of a single-mode waveguide at all three wavelengths would be preferable and likely lead to improved conversion efficiencies. Alternatively, an increased efficiency should be possible by exploring methods to reduce higher-order mode excitation, such as more careful mode matching between the Gaussian free-space mode and the elliptical waveguide fundamental mode. Our efficiency is comparable to the highest values achieved in frequency conversion experiments performed to date \cite{Roussev:04, Albota:04, Langrock:05, Pelc:11, Kambs:16}. 

\subsubsection{Noise at the single-photon level.}

Any way in which the strong pump laser field can introduce photons directly at the output telecom wavelength introduces  noise that can dominate the single-photon QFC signal. We call the rate of such noise the noise photon rate (NPR) to distinguish it from the intrinsic detector dark count rate (DCR).

At 200~mW, the pump photon flux in the waveguide is vast ($~2\times 10^{18}$ Hz) such that even extremely weak  processes in the waveguide, through which pump photons are converted directly to telecom, can overwhelm the output. 
The key process through which this happens on propagation through the crystal is anti-Stokes Raman scattering, where pump photons receive energy from phonons in the crystal. Anti-stokes Raman spectra for similar ridge waveguides to ours are presented in \cite{Zaske:11} and show that the large spectral separation between our 1902~nm pump and 1550 nm target is far from any Raman resonance peaks. Nevertheless, significant photon noise has been observed at spectral separations well beyond that expected by theory \cite{Pelc:11}, covering even the large spectral separation in our experiments. 

We study the noise photons at the waveguide output for 200~mW input pump light only, using the single-photon analysis path at the waveguide output in Figure \ref{figure1}. That analysis path consists of various removable filters and finally a single-mode-fiber-coupled, free-running InGaAs solid-state single photon detector (IDQuantique ID230 NIR). The detector is operated in a regime with the highest ratio of efficiency (10\%, as specified) to dark counts ($1.8\pm 0.1$ counts/s), achieved with a deadtime of 20 $\mu$s. 

\begin{figure}[ht]
	\centering
  \includegraphics[width=0.45\textwidth, angle=0]{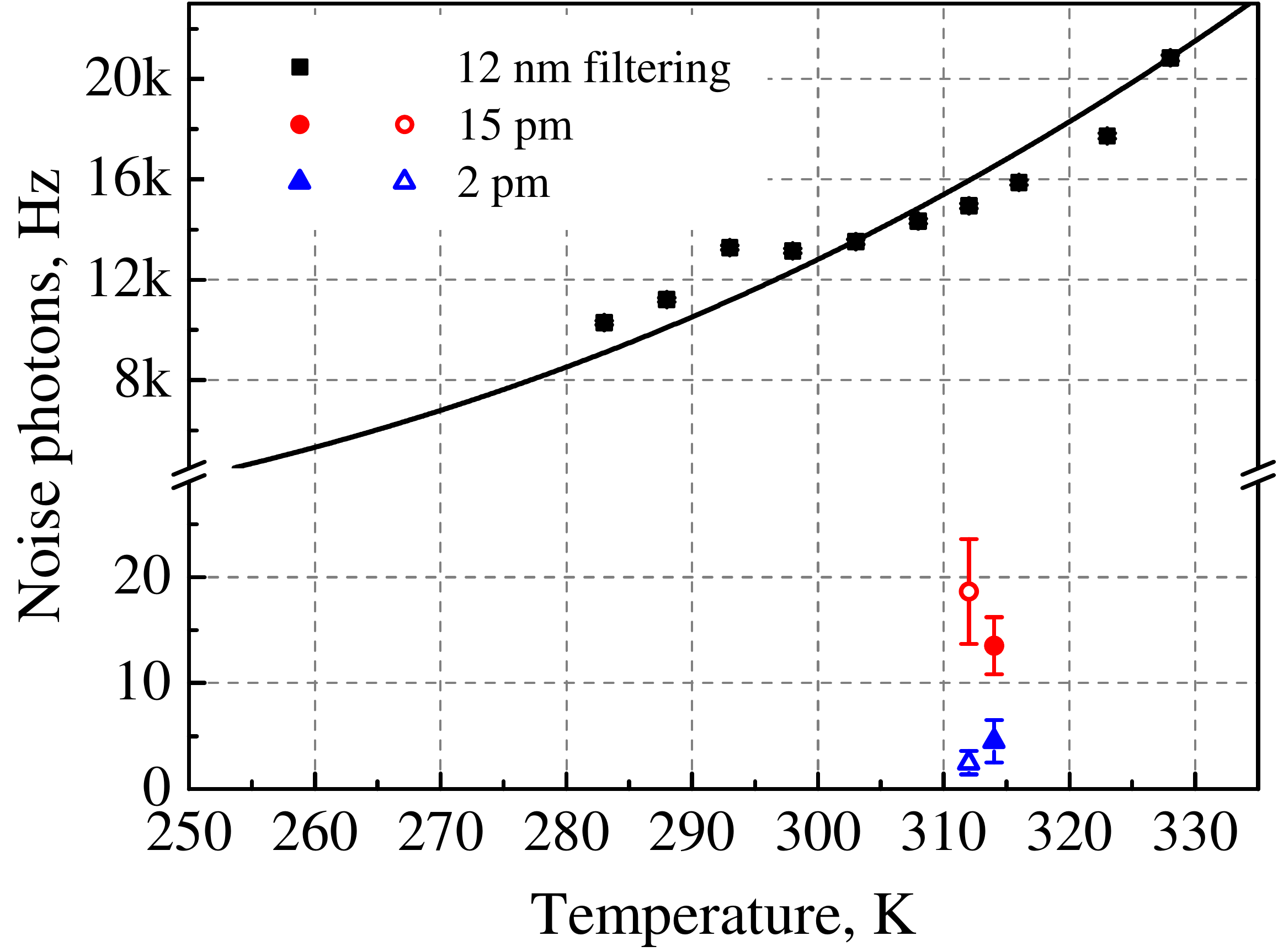}
	\caption{\textbf{1550 nm noise photons produced directly by interaction of the strong pump with the waveguide}. Plotted values are ten times the count rate of the single photon telecom detector, for 200~mW of pump light input into the waveguide alone (Figure 1). The factor of ten accounts for the 10 \% detector efficiency, yielding the total noise photons within the filtering bandwidth before detection. Rates are shown for three different filtering bandwidths centred at 1550 nm, as labelled, and for different temperatures of the waveguide. Filled points represent experimental data. Empty points show the count rate expected for the corresponding filtering bandwidth, when starting with the observed values for 12~nm bandwidth and assuming that the noise spectrum is white (points slightly shifted to the left for clarity). 
12 nm (squares) and 15 pm (circles) filtering bandwidth values are normalised to total transmission efficiency of 2 pm bandwidth filtering stage (triangles), i.~e. multiplied by factors 0.98 and 0.82 respectively, to allow for direct comparison. Solid line shows the fit by Boltzmann distribution (see the main text).}
	\label{figure3}
\end{figure}

For room temperature operation and a 12 nm filtering bandwidth, we observe a total detector click rate of 1.4 kHz, which is completely dominated by noise photons. When accounting for our detector efficiency of 0.1, that noise rate corresponds to 14 kHz of photon noise in the filtering bandwidth before detection (Fig.~\ref{figure3}). The NPR is seen to reduce with crystal temperature $T$, as one would expect for anti-Stokes Raman scattering, approximately following the phonon occupation number given by the Boltzmann distribution $\mathrm{NPR}= Ae^{-\frac{\hbar\Delta\omega}{k}\cdot\frac{1}{T}}$, where $\Delta\omega$ is pump-target frequency difference; $\hbar$, $k$ are Planck and Boltzmann constants, $A$ is a fitting parameter. By extrapolation of the theoretical fit, operation at -50~$^{\circ}$C could provide a total noise reduction by a factor of 9. This offers a way to reduce photon noise, without the need for narrowband filtering.   

The NPR at the quasi-phase matching temperature is seen to reduce in proportion to the filtering bandwidth (Figure~\ref{figure3}), consistent with the noise source being broadband and white. Using our narrowest filtering bandwidth of 2 pm at 1550 nm (250~MHz bandwidth, transmission $0.26\pm{0.01}$ at 1550 nm), the NPR before detection is reduced to 4 $\pm{2}$ Hz. 
Note, in Section \ref{scheme2} we employ a few picometer filtering stage with a greatly improved transmission at 1550 nm. 

\subsubsection{Efficiency and signal-to-noise ratio, at the single photon level.}

In order to determine a signal-to-noise ratio (SNR) one has to determine an appropriate signal: the rate of 854 nm photons one can expect to be available for converting in experiments involving a trapped Ca$^+$ ion. We consider the case where 854 nm photons are to be generated on-demand by (e.g. triggered by a laser pulse), and are entangled with,  the Ca$^+$. In \cite{Stute:2012fk}, the repetition rate for attempting to generate photons was $\sim$1~kHz, while photons were actually collected into optical fiber at a rate of $\sim$100~Hz. In future CQED ion trap systems, that use state-of-the-art mirror coatings with losses of only a few parts per million \cite{PhysRevLett.67.1727}  (to enable higher collection efficiencies), and faster cooling schemes for state reinitialisation \cite{PhysRevA.93.053401} (to enable higher repetition rates), it is feasible that triggered 854 nm photons could be collected at a rate of 10~kHz from the ion. 

\begin{figure}[ht]
	\centering
  \includegraphics[width=0.5\textwidth, angle=0]{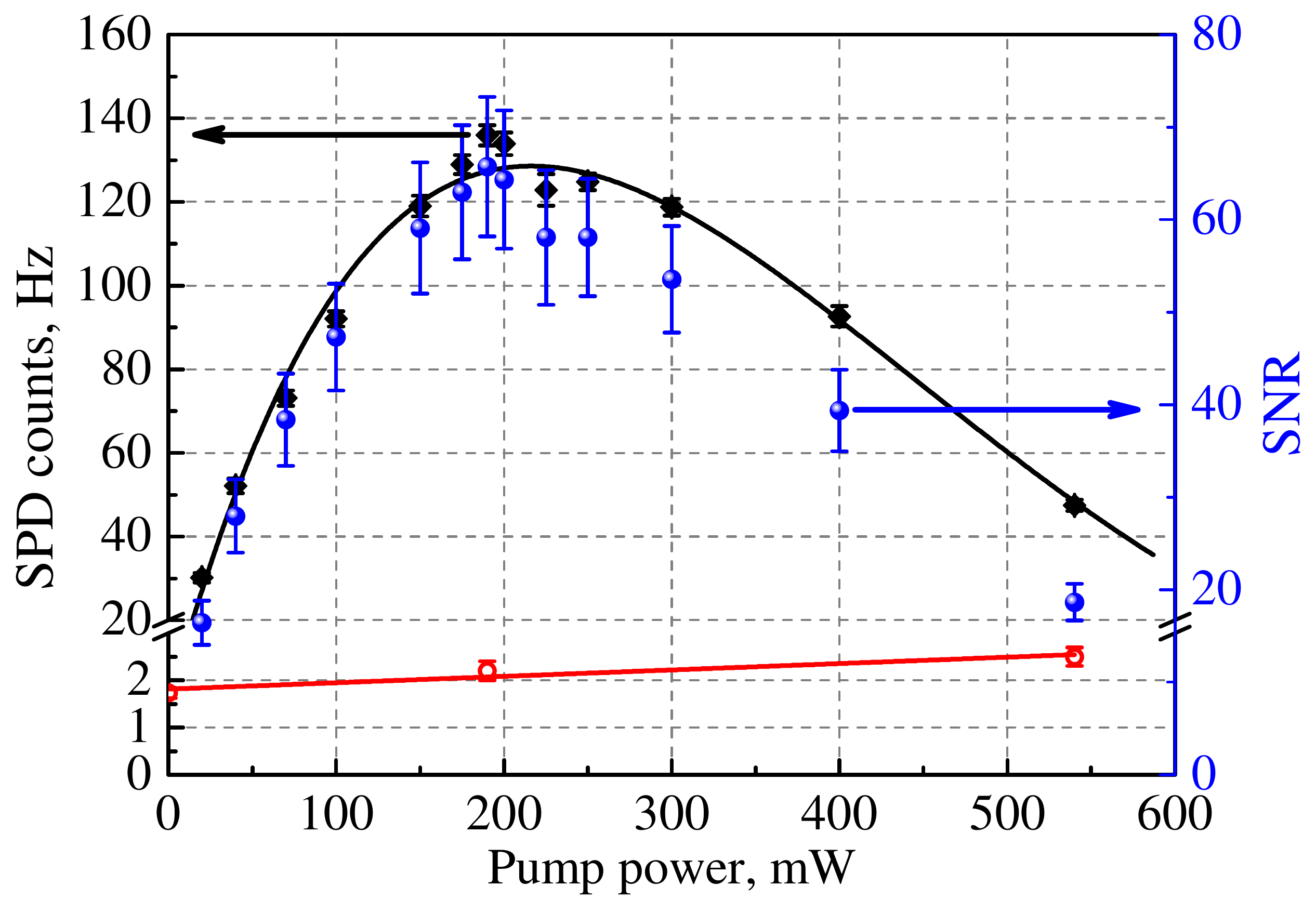}
	\caption{\textbf{Conversion efficiency, of single-photon-level 854 nm laser light to 1550 nm, as a function of pump power.} Black diamonds: telecom single photon detector count rate for an average of 10 kHz 854 nm photons at the waveguide input (signal). Black solid line: fit by function $\eta = A\cdot\sin^2{(\sqrt{\eta_{nor}P_p}L)}$ with fitting parameters $A = 128.6$, $\eta_{nor} = 0.50 \mathrm{W^{-1}cm^{-2}}$. Red circles: single photon detector count rate when the 854 nm input is blocked (noise), Red straight line: linear fit of noise. Blue spheres: signal to noise ratio. All errors bars represent one standard deviation and derive from Poissonian photon counting statistics.}
	\label{figure4}
\end{figure}

Using calibrated neutral density filters, the input 854 nm light power in our setup is attenuated to a value corresponding to an average photon rate of 10~kHz (2 fW) before the input aspheric lens, to replicate a future trapped-ion source. Figure~\ref{figure4} presents the photon count rate of the 1550 nm detector as a function of pump power and for our 2 pm bandwidth telecom filter.  At the peak conversion efficiency (again at around 200~mW pump) 136 $\pm{3}$  Hz counts are recorded, corresponding to a total detected conversion efficiency of 0.0136$\pm{0.0004}$.  
When removing the 0.10 efficiency of our detector this corresponds to a photon in/out conversion efficiency of $\eta_{out/in}=0.136$: the probability that an incoming 854 nm photon is converted to a single-mode-fiber-coupled 1550 nm photon that has passed the filtering stage. This result is consistent with our classical light measurements, leading to an external waveguide conversion efficiency of $0.62\pm{0.03}$ when removing the filtering losses. 

Since the detected total noise rate is only a few Hz, the SNR profile shown in Figure \ref{figure4} closely follows the conversion efficiency curve, peaking at  66$\pm{6}$ at 200~mW. 
A SNR $>$ 1 would therefore be possible for an input  854 nm photon rate 60 times lower than was used ($\sim$ 160 Hz) bringing QFC experiments using 854 nm photons from an ion within reach of existing experimental systems.
 \\

\section{Polarisation-preserving conversion} \label{scheme2}

For applications in quantum networking, the conversion process should preserve any photonic degree of freedom used to encode quantum information. While the single-waveguide conversion scheme presented in Figure~\ref{figure1} is suitable for e.g. time-bin encoded photonic qubits, only one polarisation component is converted. Polarisation qubits are appealing as they are straightforward to manipulate, analyse and can be preserved through long optical fibers \cite{1367-2630-11-4-045013}. Furthermore, polarisation entanglement between ion and 854 nm photon, as well as state mapping to photonic polarisation have been achieved experimentally \cite{Stute:2012fk,statetransfer}. 

\subsection{Experimental details}

\begin{figure}[t]
	\centering
  \includegraphics[width=0.45\textwidth, angle=0]{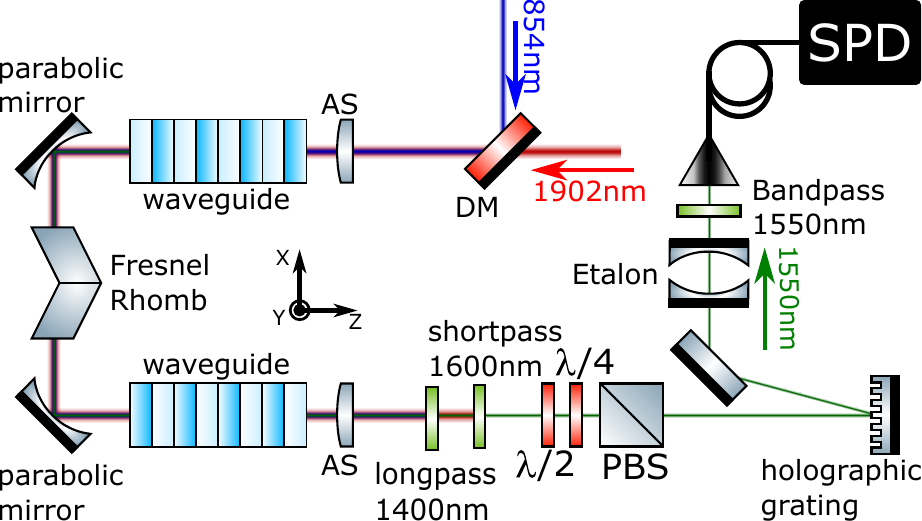}
	\caption{\textbf{Setup for polarisation-preserving frequency conversion from 854 nm to 1550 nm}. Beam paths for in-couping fields before dichroic mirror (DM) are the same as in fig. \ref{figure1}. Gold parabolic mirrors are used for simultaneous focusing/collimation of all fields (F =15 mm, Thorlabs MPD00M9); holographic grating -- volume Bragg grating, reflection bandwidth 0.2nm (25GHz), 95\% reflection (OptiGrate); bandpass 1550 nm filter has 12nm bandwidth and 95\% transmission (Thorlabs FBH-1550). Shortpass, longpass and etalon filters are the same as used in section \ref{scheme1}. Aspheres and parabolic mirrors are placed on XYZ translation stages. The first waveguide rests on an X translation stage and the second on XZ stages. Each crystal is independently temperature stabilized for optimal phase matching.}
	\label{figure5}
\end{figure}

Our polarisation-preserving conversion scheme, which employs two independent waveguide crystals in series  (Figure~\ref{figure5}), is now briefly summarised. The conversion process in each waveguide is phase matched when all three optical fields have the same linear polarisation (vertical $\ket{V}$); the orthogonal  polarisation (horizontal, $\ket{H}$) is supported in the guide but remains unconverted. Consider an arbitrary input 854 nm single-photon polarisation state $\alpha \ket{H_{854}} + \beta \ket{V_{854}}$ and input classical pump polarisation state $ \delta \ket{H_{pump}} + \gamma \ket{V_{pump}}$. The first waveguide in Figure~\ref{figure5} converts the component $\beta \ket{V_{854}}$ to $\beta \ket{V_{1550}}$ with an efficiency that depends on $\gamma$. Next, a Fresnel Rhomb (equivalent to a broadband waveplate) performs a flip operation on the polarisation of all three optical fields, converting e.g. $\ket{H}$ to $\ket{V}$ and vica versa. The second waveguide then converts the remaining 854 nm polarisation component, with an efficiency that depends on $\delta$. In the case of balanced  conversion efficiency for each polarisation, and after renormalisation,  the output telecom photon polarisation state is $\alpha \ket{V_{1550}} + \beta \ket{H_{1550}}$. In this setup we employ a filtering stage with a 4 pm bandwidth and transmission of 73 \% for fiber-coupled telecom output photons, afforded via a volume holographic grating.   
Note that interferometric path length stability between the waveguides is not required as all optical fields follow the same path. \\

\subsection{Results}

\subsubsection{Efficiency and noise, at the single-photon level.} 
To characterise the device, 854 nm laser light is injected with an average of 10 kHz photons to mimic the ion source. The power in each polarisation component of the pump is set to both maximise and balance the conversion efficiency for each polarisation component of the 854 nm light  (approximately 200~mW of the relevant pump polarisation component in-coupled into each waveguide).  
When injecting V (H) polarised 854 nm light, telecom photons are detected at a rate of 304~Hz $\pm{6}$ (302~Hz $\pm{6}$) without polarisation analysis. This corresponds to a total efficiency of 0.304 $\pm{0.006}$ (0.302 $\pm{0.006}$) when removing detector inefficiency. 
That is, the probability of obtaining a fiber-coupled telecom output photon is $\sim$0.30 for both polarisations. 
The total measured (noise) count rate, when blocking the input 854 nm light, is 7.6 Hz $\pm{0.4}$, yielding a device signal to noise ratio of $\sim$ 40. That noise count rate corresponds to a fiber-coupled telecom NPR of 58 $\pm{4}$ Hz, when subtracting dark counts and removing detector inefficiency.

\subsubsection{Polarsation-preservation.} 
To determine the extent to which the polarisation is preserved, as it is converted from 854 nm to 1550 nm, we perform process tomography \cite{MikeIke}. Specifically, the 6 standard basis states for polarisation qubits are injected (horizontal, vertical, diagonal, antidiagonal, right- and left-circular). For each input, sufficient measurements are performed to reconstruct the output 1550 nm polarisation state, via single-photon counting. From these measurements, the process matrix $\chi_{ij}$ is reconstructed via an optimisation to find the most likely process to have generated the data (maximum likelihood estimation). The process matrix describes the process $\varepsilon$ applied to any input polarisation density matrix $\rho$ via $\varepsilon(\rho)=\sum_{ij}\chi_{ij}O_{i} \rho O_{j}^{\dagger}$, where $O_{k}$ are a set of operators which form a basis for the set of operators on the polarisation qubit state space. We choose the basis $\{O_{1},O_{2},O_{3},O_{4}\}=\{I, Z, X, -iY\}$, corresponding to the identity and standard Pauli operators, respectively. 
To quantify polarisation-preservation, we maximise the $\chi_{11}$ (identity) element of the reconstructed process matrix when allowing for arbitrary unitary operations (via a simple numerical search), obtaining a value of 0.93 $\pm{0.01}$: this is the minimum fidelity with which any input polarisation state is translated through the device and detected, up to a fixed known unitary operation, and is more than sufficient to preserve polarisation entanglement between ion and photon, as discussed in the following section. We find that a fidelity of 0.95 $\pm{0.01}$ would be achieved when accounting for the measured NPR and DCR. The remaining infidelity is attributed to errors in the angular settings of waveplates by less than a degree.

\section{Future applications: long-distance transmission.} \label{long-distance}

 Converting photons from 854 nm to 1550 nm offers a reduction in attenuation in optical fiber from 3 dB/km to 0.2 dB/km, respectively. When accounting for the finite 30\% efficiency, one finds that our conversion system offers an improved rate of photon transmission for all fiber lengths greater than 1.9 km. The advantage becomes profound for longer distances: over 50 km (100 km) of fiber, the transmission probability using our device would be 3\% (0.3\%), compared to $1 \times 10^{-13}$\% ($1 \times 10^{-28}$\%) at 854 nm. 

Using photons to distribute entanglement between remote network nodes is an important task in quantum networking. We are interested therefore in assessing the distance over which our conversion and detection setup could allow for entanglement to be detected between a photon and an ion. To answer this question, we consider the case where the ion emits 854 nm photons on demand, at a rate of 10 kHz and in the maximally entangled state $(\ket{g,H_{854 nm}}+\ket{e,V_{854 nm}})/\sqrt{2}$, where $g$ ($e$) are orthogonal electronic states of the ion \cite{Stute:2012fk}. Next, we apply a modified version of the process matrix describing our converter to the photon part of the entangled state. The process matrix is modified in the sense that it is reconstructed from experimental data after subtracting detector dark counts, leaving imperfection due to photon noise. Photon noise is treated separately from dark counts at this point, since the former attenuate at the same rate as photons from the ion through a subsequent optical fiber, while detector dark counts do not. 
Finally, we apply a second process to the photon state which accounts for the 30\% conversion efficiency,  transmission probability through optical fiber of length L, detection efficiency (10\%) and dark counts at 1.8 Hz (modelled by a depolarisation channel weighted in proportion to transmitted signal). Entanglement in the final ion-photon state is quantified by the negativity \cite{PhysRevA.65.032314}, although other measures give equivalent results. The result is that entanglement between ion and photon is present up until 84 km of optical fiber. Beyond this distance, detector dark counts overwhelm the (imperfect) entanglement in the converted ion-photon state. For reference, after 84~km, telecom photons from the conversion process should be detected at a rate of 6.3 Hz, compared with the dark counts of 1.8 Hz, yielding a SNR of (6.3+1.8)/1.8 = 4.5 (photon noise is negligible).  

The achievable distance for ion-photon entanglement in our setup could be significantly extended by considering only those detection events that occur when the ion could have generated a photon. A generation rate of 10 kHz for photons from the ion allows for 100 $\mu$s per photon, of which less than 20 $\mu$s consists of the photon wave packet itself \cite{Stute:2012fk}. The remaining 4/5 of the time is allocated for reinitialising the ion after each attempt, during which time  counts at the detector can be ignored, allowing the dark counts to be reduced to 1.8/5 = 0.4 Hz.  With such a reduced dark count rate, the maximum achievable distance for the observation of ion-photon entanglement in our setup is 122 km. 

Note that, under ambient conditions the polarisation in a 25 km telecom fiber spool is known to be passively stable over timescales of several minutes \cite{1367-2630-11-4-045013}. To correct for polarisation rotations in long fibers between remote locations, one could consider periodic calibrations with classical fields, or even calibrating continuously using classical fields in parallel with the quantum signal: exploiting the narrowband nature of trapped-ion networking photons for filtering.

Transmission of light-matter networking photons through optical fibers many tens of kilometres long poses an interesting problem: the photon travel time eventually becomes longer than the minimum time between photons set by the generation rate. In our example we considered a 10 kHz generation rate, corresponding to a photon every 100 $\mu$s, whereas the travel time over 84 km (100 km) is 420 $\mu$s (610 $\mu$s). With a single ionic-qubit in a trap, one must wait to see if a generated photon is detected after the fiber, before attempting to generate another photon, or entanglement with the first photon will typically be lost. For single ion qubit experiments, the maximum generation rate is therefore limited by the photon travel time. With multiple ionic qubits in a trap, however, one could envisage different ways to overcome the photon travel time limit. For example, after generating a photon with which it is entangled, the state of the ion could be swapped into one of a collection of co-trapped ions, freeing up the original ion to generate a new photon without destroying entanglement with the first photon. First-step experiments in this direction have recently been performed \cite{monroeModular}.

\section{Conclusion and Discussion} \label{conc}

We have demonstrated a polarisation-preserving photonic interface between the 854 nm transition in trapped Ca$^+$  and the 1550 nm telecom C band. 
A total photon in / fiber-coupled telecom photon out efficiency of 30\% was achieved, for a free-running photon noise rate of $\sim$ 60 Hz. This highly efficient and low noise converter will enable telecom conversion using existing trapped-ion systems with a SNR $>1$. In combination with near-future trapped-ion systems, our converter allows for the distribution of ion-photon entanglement over more than 100~km of optical fiber, opening up the possibility of building large-scale light-matter quantum networks.  

In principle, our device should function equally as well in reverse: allowing 1550 nm photons to be converted to 854 nm  via sum-frequency generation with the pump laser. 

For the experiments presented in this work it was not necessary to stabilise the frequency of the pump laser, since the conversion process bandwidth (several tens of GHz) is large compared to the frequency drift rate of the pump laser (few hundred MHz per hour). However, the spectral properties of the pump laser will be transferred onto the converted photon: whether that is important depends on the particular experiment that one wishes to perform with frequency converted photons. For example, for the observation of ion-photon polarisation entanglement, frequency broadening of the converted photon should play no role. For entanglement swapping between remote nodes via one- or two-photon detection \cite{RevModPhys.82.1209}, frequency distinguishability between photons becomes important.  
For schemes where remote nodes absorb photons that have been frequency converted (e.g. conversion to telecom then back to 854 nm for absorption by a remote ion), special care may need to be taken to stabilise the pump laser frequency to minimise the spectral footprint on the photons. One could also consider using the temporal and spectral properties of the pump as a way to coherently modify those properties of the converted photons, to overcome bandwidth mismatches between remote quantum matter. 

During the preparation of this manuscript we became aware of complimentary work to ours, in which polarisation-preserving conversion from 854 nm to 1310~nm (Telecom O band) was achieved using a single ridge waveguide scheme in a cavity \cite{Lenhard:17}.

\section{Acknowledgements}

We thank Carsten Langrock, Christian Roos and Petar Jurcevic for comments on the manuscript and the staff at IQOQI for technical and administrative support. We thank Rainer Blatt for providing the laboratory space, environment and group support in which to develop our work. This work was supported by the START prize of the Austrian FWF project Y 849-N20, the Army Research Laboratory Center for Distributed Quantum Information via the project SciNet and the Institute for Quantum Optics and Quantum Information. 


\end{document}